\documentclass[%
  preprint,
 amsmath,amssymb,
]{revtex4-2}

\usepackage{dcolumn} 
\usepackage{bm}      
\usepackage{setspace}
\usepackage{amsfonts}
\usepackage{braket}

\usepackage{graphicx}

\newcommand{\del}[3] {\frac{\partial^{#3} #1}{\partial #2^{#3}}}
\newcommand{\dev}[3]{\frac{\text{d}^{#3} #1}{\text{d}#2^{#3}}}

\newcommand{\pdel}[3]{{\partial^{#3} #1}/{\partial #2^{#3}}}

\newcommand{\ee}{{\bm e}}

\newcommand{\mm}{{\bm m}}
\newcommand{\pp}{{\bm p}}
\newcommand{\rr}{{\bm r}}

\newcommand{\tm}{{\yy_{tag}}}
\newcommand{\uu}{{\bm u}}

\newcommand{\ww}{{\bm w}}
\newcommand{\xx}{{\bm x}}
\newcommand{\yy}{{\bm y}}
\newcommand{\zz}{{\bm z}}

\newcommand{\BB}{{\bm B}}

\newcommand{\RR}{{\mathbb R}}

\newcommand{\FF}{\mbox{\boldmath $F$}}

\newcommand{\GG}{\mbox{\boldmath $G$}}
\newcommand{\tr}{\mathrm{T}}

\usepackage{amsmath}	
\begin{document}

\preprint{APS/123-QED}

\title{
Blending optimal control and biologically plausible learning for noise-robust physical neural networks
%
}

\author{
Satoshi Sunada$^{1}$,
Tomoaki Niiyama$^{1}$,
Kazutaka Kanno$^{2}$, Rin Nogami$^{2}$, 
Andr\'e R\"ohm$^{3}$, Takato Awano$^{1}$, and Atsushi Uchida$^{2}$ 
}

\affiliation{%
%
${}^1$Faculty of Mechanical Engineering, Institute of Science and
Engineering, Kanazawa University,
Kakuma-machi, Kanazawa, Ishikawa 920-1192, Japan \\
${}^2$Department of Information and Computer Sciences, Saitama
 University, 255 Shimo-Okubo, Sakura-ku, Saitama City, Saitama,
 338-8570, Japan\\
${}^3$Department of Information Physics and Computing, Graduate School of Information Science and Technology, The University of Tokyo, 7-3-1 Hongo, Bunkyo-ku, Tokyo 113-8656, Japan\\
}%

\date{\today}

\begin{abstract}
The rapidly increasing computational demands for artificial intelligence (AI) have spurred the exploration of computing principles beyond conventional digital computers. Physical neural networks (PNNs) offer efficient neuromorphic information processing by harnessing the innate computational power of physical processes; however, training their weight parameters is computationally expensive. We propose a training approach for substantially reducing this training cost. Our training approach merges an optimal control method for continuous-time dynamical systems with a biologically plausible training method---direct feedback alignment. In addition to the reduction of training time, this approach achieves robust processing even under measurement errors and noise without requiring detailed system information. The effectiveness was numerically and experimentally verified in an optoelectronic delay system. Our approach significantly extends the range of physical systems practically usable as PNNs.
\end{abstract}

\maketitle

{\it Introduction.---}The availability of vast amounts of data and computing power has spurred rapid development of deep learning. However, the increasing power demands for deep learning have outpaced the advancements in digital electronics \cite{Mehonic:2022wi}. This situation has motivated the development of novel computing technologies that differ significantly from the existing von Neumann computers \cite{Furber_2016, Nahmias:18,RevModPhys.91.045002,Merolla668,Grollier:2020aa,Markovic:2020aa,9049105,Shastri:2021aa,Nakajima:2022aa,Science.abq8271_Delocalized,Sciadv.aay6946_wavephysics,PhysRevResearch.3.033243,Zhang:2020aa,TANAKA2019100,Wright:2022aa,doi:10.1126/science.adi8474}. 

Utilizing physical systems enables efficient neuromorphic information processing by leveraging the intrinsic properties of the systems themselves \cite{Wright:2022aa,doi:10.1126/science.adi8474,momeni2024trainingphysicalneuralnetworks}. One example that has garnered significant attention in physical sciences is reservoir computing (RC) \cite{TANAKA2019100,Jaeger78,Maass:2002,VERSTRAETEN2007391,Appeltant:2011ab,Liang:2024aa} or extreme learning machines (ELMs) \cite{HUANG2006489,Ortin:2015aa}, which can utilize physical dynamics for information processing, making it powerful for processing time-dependent or static data. However, RC and ELMs are limited in tunability as they are trained only at the readout weight parameters, which restricts their information processing capabilities.

\begin{figure}[htbp]
\centering\includegraphics[width=8.6cm]{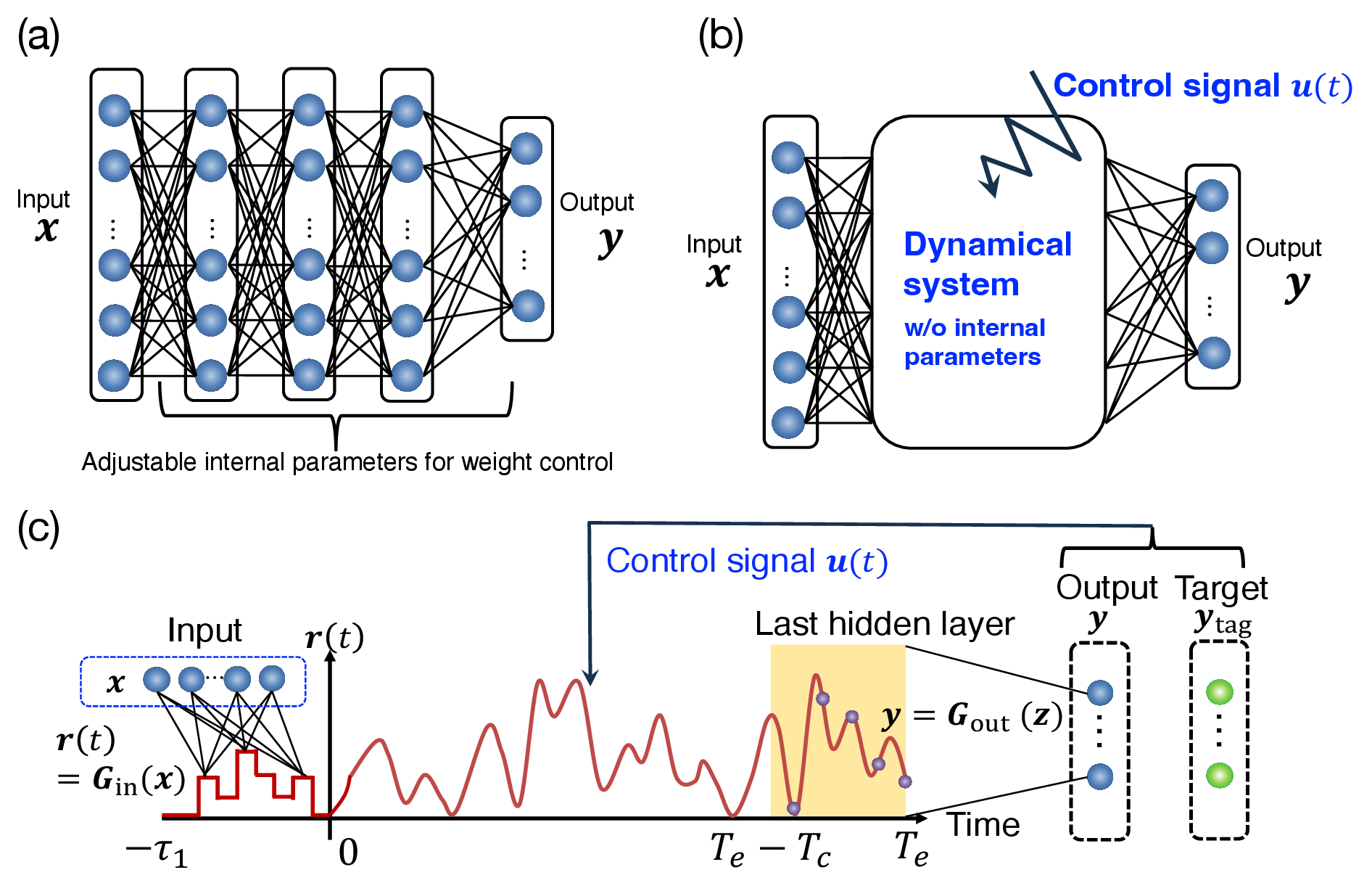}
\caption{\label{fig_schematic}
(a) Schematic of physical neural networks (PNNs) with adjustable internal parameters for weight control. 
(b) Physical system driven by external control signals.
(c) Proposed approach based on external control. 
}
\end{figure}

Deep neural networks (DNNs) offer high expressivity owing to its layer-to-layer information propagation, enabling advanced processing for practical tasks. A straightforward approach to physically implementing DNNs is using a physical system with adjustable internal elements for weight control \cite{PhysRevApplied.18.014040,Shen:2017aa,Feldmann:2021aa}. (See Fig.~\ref{fig_schematic}(a)). However, designing such a physical system with numerous adjustable elements remains challenging for large-scale computations. 

This study considers an alternative approach that involves driving physical systems externally [Fig.~\ref{fig_schematic}(b)], eliminating the need for adjusting internal elements. It offers the following features:  
First, it enables advanced learning even in physical systems without tuning elements, using only a few external controllers \cite{PhysRevApplied.15.034092}. Second, the approach offers a high expressivity for information processing of continuous-time dynamical systems \cite{NEURIPS2018_69386f6b}, unlike DNNs that are associated with discrete-time dynamical systems. Third, concepts, such as neurons and synapses, do not appear explicitly, allowing for a direct correlation between the dynamical process and DNN-like information processing. This provides insights into the learning mechanism from the perspective of optimal control of dynamical systems.  

However, optimal control methods for dynamical systems encounter a common challenge shared with the backpropagation method utilized in DNNs.
The parameter updates are based on error backpropagation, necessitating a substantial amount of computational energy and time for training. 
Conversely, biological systems achieve efficient learning in noisy and dynamic environments without relying on error backpropagation \cite{Jaeger:2023uw}. 

This study proposes a training approach to significantly reduce the computational training cost by incorporating a biologically plausible learning strategy known as direct feedback alignment (DFA) \cite{nkland2016direct} into an optimal control method for dynamical systems. This approach is based on external driving and can be applied to various dynamical systems without requiring numerous adjustable internal parameters, unlike previous approaches \cite{Wright:2022aa,Nakajima:2022aa,doi:10.1126/science.adi8474}; therefore, it is versatile and useful for extracting the computational capabilities of physical systems. Moreover, the proposed approach enables highly robust processing amid noise and yields excellent performance. 
This approach can unlock the computational potential of numerous physical platforms.

{\it Optimal control-based model.---}The primary objective is to identify a function that can map an input $\xx \in \RR^{N_x}$ onto its corresponding target $\tm \in \RR^{L}$ for a given training dataset, where $N_x$ and $L$ denote the degree of freedom of input data and target information, respectively. To express the input-target relationship, this study employs a continuous-time dynamical system with time delays, which is governed by 
\begin{align}
\dev{\rr(t)}{t}{}  = \FF\left[
\rr(t),\rr(t-\tau_D),\uu(t)
\right], \label{eq_system} 
\end{align}
where $\rr(t) \in \RR^{N_r}$ represents the state vector at time $t$ and $\uu(t) \in \RR^{N_u}$ represents $N_u$ control signals. $\tau_D$ denotes the delay time. The dynamical system is driven by external control signals $\uu(t)$, such that output $\yy$ corresponds to a target vector $\tm$ for a given input $\xx$. The schematic is shown in Fig.~\ref{fig_schematic}(c). Based on the correspondence between DNNs and dynamical systems \cite{PhysRevApplied.15.034092}, an input $\xx$ is encoded in an initial state as $\rr(t) = \GG_{in}(\xx)$ for $-\tau_D \le t \le 0$. The input information is transformed by the time evolution (forward propagation) of the dynamical system up to the end time $t = T_e$. Let $T_c$ be the time length of the output layer, respectively. $\yy$ is expressed as a function of $\rr(t;\xx)$ for $T_e - T_c < t \le T_e$ as $\yy = \GG_{out}\left(\zz\right)$, where $\zz = \int_{T_e-T_c}^{T_e}\ww(t)\rr(t) dt$ and $\ww(t) \in \RR^{L\times N_r}$ represents an output weight function. The learning objective is to identify optimal control $\uu^{*}(t)$ and $\ww^{*}(t)$ that minimize the loss function $L(\yy,\tm)$ for a given training dataset.

The output weight function $\ww(t)$ can easily be updated by utilizing the gradient of $L$ with respect to $\ww(t)$ as $\ww(t) \rightarrow \ww(t) - \eta_w \pdel{L}{\ww}{}$, where $\eta_w$ represents the learning rate. This does not incur significant computational costs. Meanwhile, $\uu(t)$ can be trained through the adjoint method \cite{kirk2004optimal} developed within the realm of optimal control theory, as shown in Supplementary Material (SM) \cite{Supplementary}. The update of $\uu(t)$ is described by the adjoint vector $\pp(t)$, corresponding to the back-propagated error, as $\uu(t) \rightarrow \uu(t) - \eta_u \pp^{\tr}\pdel{\FF}{\uu}{}$ \cite{Supplementary}, where $\eta_u$ denotes the learning rate.
$\pp(t)$ is determined by solving the adjoint equation backward in time from $t = T_e$ to $0$, with $\pp(T_e) = 0$,
\begin{align}
\dev{\pp^{\tr}}{t}{}
&= 
-\Theta_1(t)\del{L}{\zz}{}\ww
-\pp^{\tr}\del{\FF}{\rr}{} \nonumber \\
&-
\Theta_0(t)\pp^{\tr}(t+\tau_D)\del{\FF(t+\tau_D)}{\rr}{},
\label{eq_adeq_text}
\end{align}
where $\Theta_1$ and $\Theta_0$ represent step functions that are equal to 1 only for $T_e-T_c\le t < T_e$ and $0 \le t \le T_e-\tau_D$. 
%
%
The adjoint method corresponds to backpropagation-based training for a continuous-time dynamical system described by Eq.~(\ref{eq_system}). See SM for details \cite{Supplementary}.
However, solving Eq.~(\ref{eq_adeq_text}) requires significant computational resources for training. 

{\it DFA--adjoint.---}To address these challenges, we incorporate the DFA into the adjoint optimal control method. DFA utilizes fixed random matrices and the error at the last layer to train the weight parameters \cite{nkland2016direct,NEURIPS2020_69d1fc78}. This approach helps to overcome the vanishing gradient problem and can lead to a substantial reduction in the computation steps required for backpropagation. 
Based on the correspondence between DNNs and dynamical systems \cite{PhysRevApplied.15.034092,DBLP:journals/corr/abs-1908-10920,NIPS2016_14851003}, we define the following adjoint vector \cite{Supplementary}:
\begin{align}
 \hat{\pp}^{\tr}(t) = \ee^{\tr}\BB(t), 
\end{align}
where $\ee = \yy-\tm$ represents the error at the last layer and $\BB(t) \in \RR^{L\times N_r}$ denotes a time-dependent random matrix. 
Instead of solving Eq. (\ref{eq_adeq_text}) directly, $\pp(t)$ is replaced with $\hat{\pp}(t)$.
This approach is referred to as {\it DFA--adjoint} in this study. This enables learning of the control signal, $\uu(t)$, through the following steps:

\noindent (1) Forward propagation: The dynamical system evolves over time based on given input data $\xx$.\\
\noindent (2) $\hat{\pp}^{\tr}(t) = \ee^{\tr}\BB(t)$ is computed from the output $\yy = \GG_{out}(\zz)$. \\
\noindent (3) $\uu(t)$ and $\ww(t)$ are updated.
Steps (1)--(3) are repeated until the desirable output $\yy$ is obtained.
The DFA--adjoint method corresponds to a generalized version of the DFA method, which can be applied to continuous-time systems \cite{Supplementary}.

In the DFA--adjoint method, the selection of $\BB(t)$ is crucial for achieving optimal computational performance. We determined that the characteristic timescale of the waveform of $\BB(t)$ should closely align with that of the system dynamics. 
We set the elements of $\BB(t)$ as follows:
\begin{align}
B_{li}(t)  = \sum_{m=1}^{M_B}
A_m^{(li)}\sin
\left(
\omega_m t
+
\phi_m^{(li)}
\right), \label{eq_bb}
\end{align} 
where $A_m^{(li)}$ and $\phi_m^{(li)}$ are determined by the values drawn from a uniform distribution. $\omega_m$ distributes around the characteristic angular frequency of the system dynamics. 
See SM for details \cite{Supplementary}. 

{\it Numerical simulations.---}To assess the training performance of the DFA--adjoint method, we considered a time-delay system, governed by the following equation:
\begin{align}
\tau_L \dev{r(t)}{t}{} 
&= -\left(1 + \dfrac{\tau_L}{\tau_H} \right)r(t)
- \dfrac{1}{\tau_H}\int r(t)dt \nonumber \\
&+ 
\beta\cos^2\left(
u_1(t)r(t-\tau_D) + u_2(t)
\right)
+\tau_L\sigma\xi(t), \label{eq_delay}
\end{align}
where $u_1(t)$ and $u_2(t)$ represent external driving forces to control the dynamical system. $\tau_D$, $\beta$, $\tau_L$, and $\tau_H$ represent delay time, feedback strength, and the time constants of low- and high-pass filters, respectively. 
$\xi(t)$ denotes the Gaussian noise with zero mean and unity standard deviation, whereas $\sigma$ denotes the noise strength. The equation corresponds to a model equation for an optoelectronic delay system \cite{murphy2010complex}. 

We encode input data $\xx$ as $r(t) = G_{in}(\xx) = \mm^{\tr}(t)\xx$ for $-\tau_D \le t \le 0$ using a mask pattern $\mm(t) \in \RR^{N_x}$.
From this initial state, the system evolves over time until $T_e = 3\tau_D$. The output $\yy$ is determined from the softmax functions of $\zz$ with $T_c = \tau_D$. 

To assess the performance, we utilized the MNIST handwritten digit dataset, which comprises a training set of 60,000 images, along with a test set of 10,000 images \cite{MNIST_LeCun_url}. 
The DFA--adjoint method was employed to train control signals $u_1(t)$ and $u_2(t)$, as well as $\mm(t)$ for the MNIST dataset \cite{Supplementary}. For training, a minibatch update was executed using the Adam optimizer \cite{kingma2014adam}. 

\begin{figure}[htbp]
\centering\includegraphics[width=8.6cm]{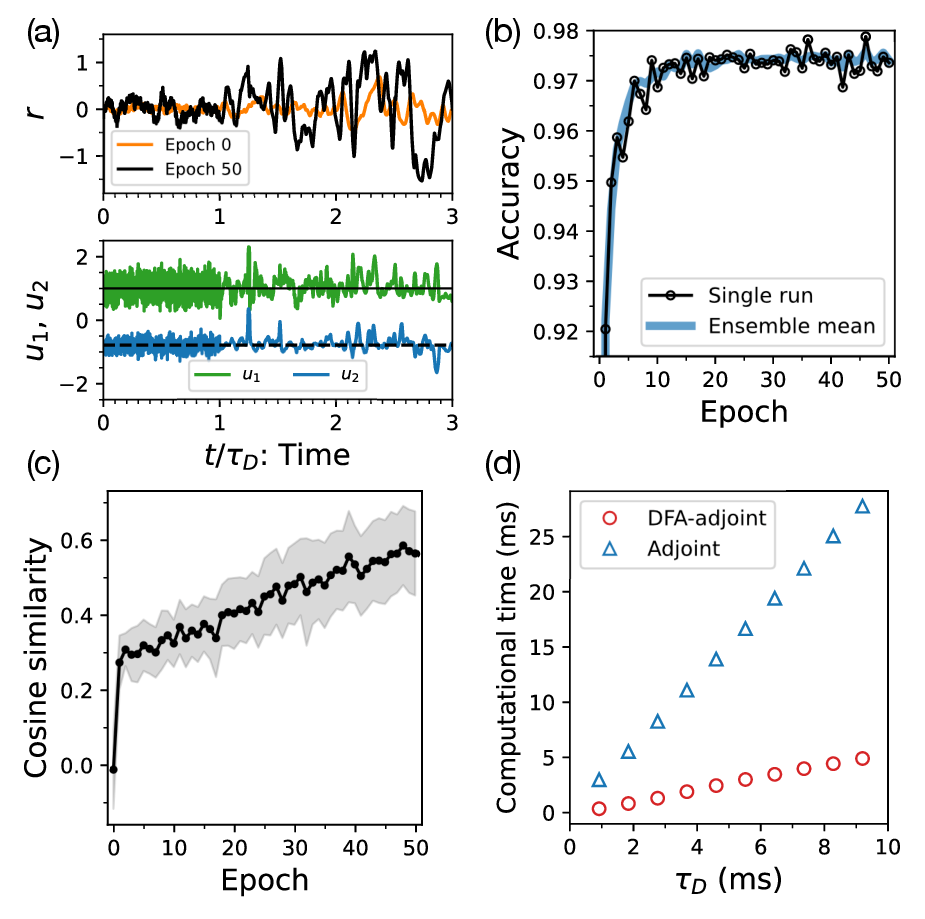}
\caption{\label{fig_result1}
(a) $r(t)$, $u_1(t)$, and $u_2(t)$ before and after training. $u_1(t)$ and $u_2(t)$ before training were represented by black solid and black dashed lines, respectively. 
(b) A typical learning curve obtained by a single run (black solid line with circles) and mean learning curve (blue solid line). 
(c) Cosine similarity in the update vectors between the DFA--adjoint and adjoint methods \cite{Supplementary}. The shadow represents the standard deviation. (d) Average computational training time per image.}
\end{figure}

Figure~\ref{fig_result1}(a) shows $r(t)$, $u_1(t)$, and $u_2(t)$ before training and after 50 training epochs. The following parameters were utilized in this simulation:
$\tau_L = 15.9$ $\mu$s, $\tau_H = 1.59$ ms, $\tau_D = 0.92$ ms, $\beta = 3.0$, and $\sigma = 0$. 
$r(t)$ exhibited changes corresponding to variations in $u_1(t)$ and $u_2(t)$. After 50 training epochs, the classification accuracy increased to 0.973 [Fig.~\ref{fig_result1}(b)].
To gain insight into the training mechanism, we measured the cosine similarity between the update vectors in the DFA--adjoint and normal adjoint (gradient-based) methods. See SM for details \cite{Supplementary}.
Figure~\ref{fig_result1}(c) shows that the similarity becomes positive, indicating that the update direction in the DFA--adjoint method tends to roughly align with that computed in the adjoint method. This trend is similar to that observed in feedback alignment applied to standard neural networks \cite{Lillicrap:2016aa}.

For further performance evaluations, we utilized the Fashion-MNIST \cite{arXiv.1708.07747} and CIFAR-10 datasets \cite{Krizhevsky09learningmultiple}
and compared the classification accuracy achieved through training with the DFA--adjoint method against results using the adjoint method and without optimal control. 
In the absence of control, the system functions as an ELM.
The results are listed in Table~\ref{tab:table1}.
The DFA--adjoint method exhibited superior classification performance compared with the control-free scenario, albeit showing a slight degradation compared with the adjoint method and previous backpropagation approach \cite{JMLR:v16:hermans}.

\begin{table}[b]
\caption{\label{tab:table1}%
Classification accuracy for MNIST, Fashion-MNIST, and CIFAR-10 datasets. Each accuracy was obtained by five runs \cite{Supplementary}.}
\begin{ruledtabular}
\begin{tabular}{lcdr}
\textrm{Methods }&
\textrm{MNIST}&
\multicolumn{1}{c}{\textrm{Fashion-MNIST}}&
\textrm{CIFAR-10}\\
\colrule
DFA--adjoint & 0.973 & 0.866 & 0.466 \\

Adjoint (Backprop.) & 0.977 & 0.881 & 0.507\\
ELM (w/o control)  & 0.869 & 0.792 & 0.275\\
\end{tabular}
\end{ruledtabular}
\end{table}

The DFA--adjoint method offered a notable advantage by eliminating the need to solve Eq.~(\ref{eq_adeq_text}) for backpropagation, significantly reducing the computational training time. Training time reduction is particularly crucial for large-scale physical systems, e.g., those with a long delay time. Figure~\ref{fig_result1}(d) shows the average computational training times per one input image for the MNIST dataset as a function of delay time $\tau_D$. The rate of the computational training time to $\tau_D$ is less than one-fifth of that for the adjoint method, making a substantial impact on training efficiency for a larger-scale system.
Further acceleration is possible when a parallel computation scheme can be used.

\begin{figure}[htbp]
\centering\includegraphics[width=8.6cm]{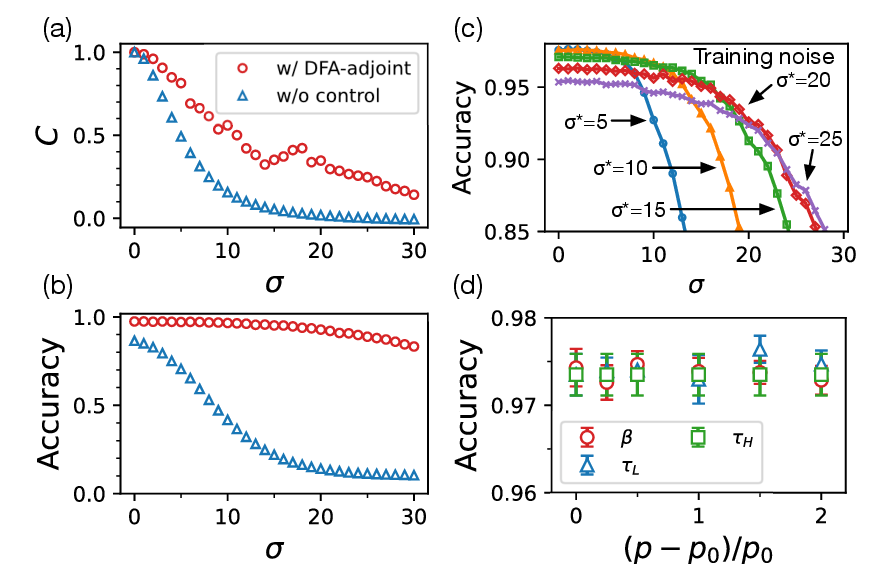}
\caption{\label{fig_result2}
(a) $C$ and (b) classification accuracy as a function of $\sigma$. 
For comparison, the results for ELM (without control) are presented.
(c) Accuracy as a function of $\sigma$ in the system trained with a different noise strength $\sigma^*$.
(d) Robustness to parameter mismatch error normalized using a true value, $(p - p_0)/p_0$, where $p$ and $p_0$ represent the parameter value used for training and a true value for $\beta$, $\tau_L$, or $\tau_H$. 
}
\end{figure}

In terms of hardware implementation, robustness against external noise is a critical requirement.
To assess robustness to noise, Gaussian noise was incorporated into Eq.~(\ref{eq_delay}), and $r(t)$ was measured starting from an initial state with various noise instances. The noise robustness was evaluated through the following cross-correlation: 
\begin{align}
C = \dfrac{1}{K}\sum_{k>k'}\dfrac{
\langle
r_k(t)r_{k'}(t)
\rangle_t
}
{\sigma_k\sigma_{k'}},
\end{align}
where $r_k(t)$ and $\sigma_k$ represent the state variable at time $t$ and the standard deviation of the $k$-th instance, respectively. $\langle \cdot \rangle_t$ and $K$ denote the time average and number of summations, respectively. 
Figures~\ref{fig_result2}(a) and \ref{fig_result2}(b) show $C$ and classification accuracy for the MNIST dataset as functions of noise strength $\sigma$, respectively. 
For this demonstration, we set $\beta = 3.0$, resulting in a chaotic behavior before training. Hence, the state variable was sensitive to noise, leading to varying outputs even for identical input data. 
This sensitivity resulted in a significant reduction in $C$. In practice, the time evolution after training in the control-free scenario led to decreasing $C$ and classification accuracy with increasing $\sigma$. 
In contrast, the DFA--adjoint method effectively mitigated the decline in $C$, thereby maintaining consistently high accuracy.
The noise robustness observed in the DFA--adjoint method can be attributed to its external driving. The system is driven by the applied control signals and compelled to adhere to these signals, rather than noise, thereby enhancing the robustness of the system dynamics.

Noise robustness is retained even when the noise strength $\sigma^*$ during the training phase differs from the noise strength $\sigma$ in the test phase. As shown in Fig.~\ref{fig_result2}(c), classification accuracy for test data can be maintained when $\sigma < \sigma^*$. This also suggests that training under strong noise perturbation makes the system more robust.

In addition to its noise robustness, the DFA--adjoint method demonstrated robustness against mismatches in system parameters. This insensitivity is crucial when utilizing an actual physical system, as the parameters may be challenging to measure accurately. 
To illustrate this point, we considered a scenario in which measurement errors existed in $\beta$, $\tau_L$, and $\tau_H$. Figure~\ref{fig_result2}(d) depicts the classification accuracy as a function of each mismatch error. 
High accuracy can be maintained even with 200 $\%$ mismatch errors. 

{\it Toward a model-free training approach.---}A drawback of the proposed training approach is its reliance on a model of $\FF$ (e.g., $\pdel{\FF}{\uu}{}$ is used for the control update). However, we found that the approach can still function effectively without an accurate model of $\FF$. High accuracy is maintained even when $\FF$ is replaced by typical activation functions, such as Gaussian or $\tanh$ functions. Additionally, we found that an appropriate design of external control can eliminate the need for prior knowledge of the physical systems in this approach. See SM for further details \cite{Supplementary}.

\begin{figure}[htbp]
\centering\includegraphics[width=8.6cm]{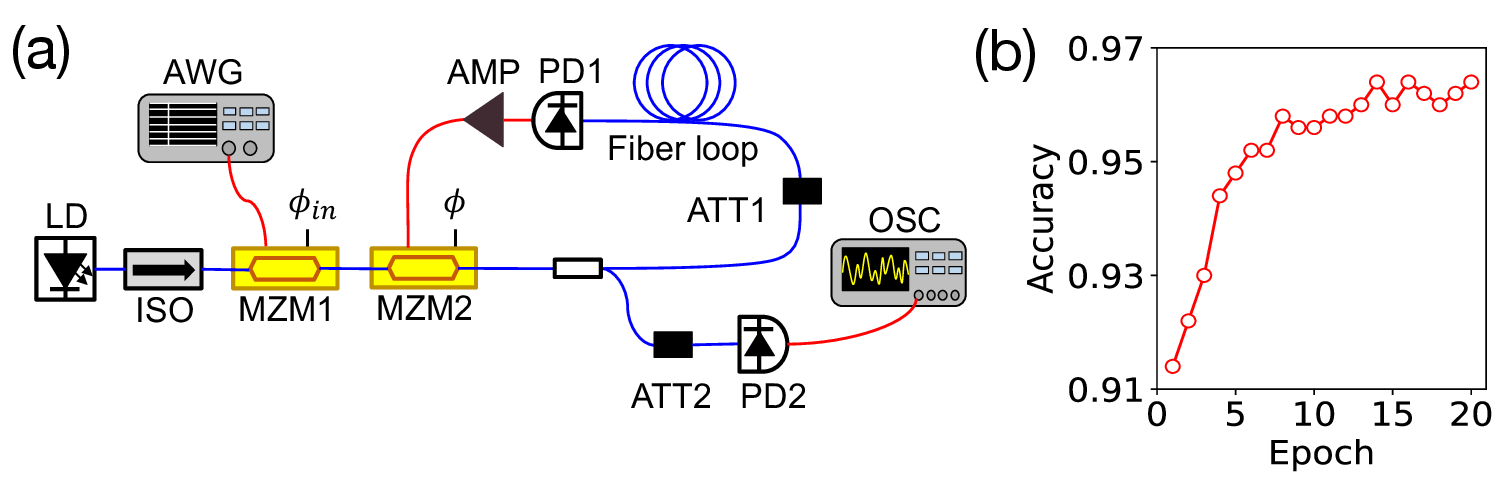}
\caption{\label{fig_exp}
(a) Experimental setup for an optoelectronic delay system. LD: semiconductor laser diode, ISO: optical isolator, MZM: Mach-Zehnder modulator, AMP: electrical amplifier, ATT: optical attenuator, PD: photodetector, OSC: digital oscilloscope, AWG: arbitrary waveform generator. (b) Learning curve. 
}
\end{figure}

{\it Experiment.---}We validated the effectiveness of the DFA--adjoint method through experiments. The experimental setup is shown in Fig.~\ref{fig_exp}(a). The system was composed of a laser diode (LD), optical fiber for delayed feedback, photodetector (PD1), and two Mach-Zehnder modulators (MZM1 and MZM2) that introduce nonlinearity of $\cos^2(\cdot)$. MZM1 was utilized for signal input and control. For $-\tau_D \le t \le 0$, the laser light was modulated by an input signal encoding data $\xx$, which was generated using an arbitrary waveform generator (AWG). For $t>0$, MZM1 was employed to control the laser light with signal $u(t)$. The optical output from MZM2 was fed back into the MZM2 after traversing a 200 m long optical fiber, with a delay time of 1.022 $\mu$s in the feedback loop. PD1 converted the feedback light to an electrical signal. The feedback gain was adjusted using an optical attenuator (ATT1) and an electrical amplifier (AMP). PD1 and AMP function as a low-pass filter with $\tau_L = 1.59$ ns and a high-pass filter with $\tau_H =  530$ ns, respectively. The bias points for MZM1 and MZM2 were both set as $-\pi/4$.

To assess the classification performance, we used the MNIST handwritten digit image dataset. 3,000 images were used for training and 500 images were used for testing. Figure~\ref{fig_exp}(b) depicts the learning curve for the system trained with the DFA--adjoint method. The classification accuracy increased to 0.964 at the 20th training epoch, which was comparable to the numerical result shown in Fig.~\ref{fig_result1}(b). 
This result shows that the proposed method can turn a real physical system with external disturbances and unknown precise parameters into a computing system.

{\it Summary.---}We proposed the DFA--adjoint method and have experimentally and numerically demonstrated its classification performance for an optoelectronic delay system. The DFA--adjoint method enables robust operation against noise and parameter mismatches. The computational training time can be significantly reduced compared to the conventional backpropagation (adjoint method). The DFA--adjoint method leverages the external control of dynamical systems; thus, it can be applied to any physical system, device, or material without adjustable elements and can unlock their computational potentials, enabling energy-efficient learning suitable for edge devices with limited resources.

This work was supported in part by JSPS KAKENHI (Grant
No. JP20H04255, No. JP22H05198, No. JP23H03467, No. JP19H00868, No. JP20K15185, No. JP22H05195), and JST, CREST Grant No. JPMJCR24R2.
%

%

%
%

\end{document}


\preprint{APS/123-QED}

\title{Supplemental Information 
}

\author{
Satoshi Sunada$^{1}$,
Tomoaki Niiyama$^{1}$,
Kazutaka Kanno$^{2}$, Rin Nogami$^{2}$, 
Andr\'e R\"ohm$^{3}$, Takato Awano$^{1}$, and Atsushi Uchida$^{2}$ 
}

\affiliation{%
%
${}^1$Faculty of Mechanical Engineering, Institute of Science and
Engineering, Kanazawa University
Kakuma-machi Kanazawa, Ishikawa 920-1192, Japan \\
${}^2$Department of Information and Computer Sciences, Saitama
 University, 255 Shimo-Okubo, Sakura-ku, Saitama City, Saitama,
 338-8570, Japan\\
${}^3$Department of Information Physics and Computing, Graduate School of Information Science and Technology, The University of Tokyo, 7-3-1 Hongo, Bunkyo-ku, Tokyo 113-8656, Japan\\
}%

\date{\today}

%
%

\maketitle

\section{Correspondence between dynamical systems and multilayered neural networks}

In preparation for discussing the DFA-adjoint method for continuous-time dynamical systems with time delay, we summarize the relationship between the adjoint optimal control method for dynamical systems and backpropagation used for training traditional multilayered neural networks (NNs).
\subsection{Backpropagation}
First, we consider an ordinary fully connected multilayered NN, 
where the output in the $n$-th layer, $\vc{z}_{n+1} \in \mathbb{R}^{M_{n+1}}$, is expressed as follows:
\begin{align}
\vc{z}_{n+1} = \FF_n(\am_n), \\
\am_n = \WW^{(n)}\vc{z}_{n}, 
\end{align}
where 
$M_{n+1}$ represents the number of nodes in the $n+1$-th layer,
$\FF_n$ represents an activation function,
and $\WW^{(n)}$ represents 
a weight matrix of the $n$th layer.
The $L$ dimensional target vector $\tm$ and the output from the last layer $\yy = \GG_\mathrm{out}(\vc{z}_N, \ww) \in \mathbb{R}^L$ determine the loss function $L(\yy, \tm)$,
where $\GG_\mathrm{out}$ represents an activation function for the output layer.
We assume that the loss function $L$ and $\GG_\mathrm{out}$ are the mean squared error (MSE) and an identity function for a regression task, respectively, whereas $L$ and $\GG_\mathrm{out}$ are the cross entropy and softmax function for a classification task, respectively. 
To minimize $L(\yy, \tm)$, the weights should be optimized 
according to the following gradient (row) vector:
\begin{align}
\dev{L}{{\WW^{(n)}}}{} 
&= 
\del{L}{\vc{z}_{n+1}}{} \dev{\vc{z}_{n+1}}{{\WW^{(n)}}}{} 
= \del{L}{\vc{z}_{n+1}}{} \del{\FF_{n}}{{\WW^{(n)}}}{}. 
\end{align}
The gradient (row) vector $\pdel{L}{\vc{z}_n}{}$ is computed backward from $\del{L}{\vc{z}_{N}}{}$ in the last layer, using the following equation recurrently:
\begin{align}
\del{L}{\vc{z}_{n}}{} = \del{L}{\vc{z}_{n+1}}{} 
\del{\vc{z}_{n+1}}{\vc{z}_{n}}{} 
= \del{L}{\vc{z}_{n+1}}{}  
\del{\FF_n}{\vc{z}_{n}}{}. \label{eq_pl}
\end{align}
Here, if $\pdel{L}{\vc{z}_{n}}{}$ is defined by $\pp^{\tr}_{n}$ as follows: 
\begin{align}
 \pp^{\tr}_{n} = \del{L}{\vc{z}_{n}}{},  
\end{align}
where $^{\tr}$ represents transpose, 
Eq.~(\ref{eq_pl}) is rewritten as follows:
\begin{align}
\pp^{\tr}_{n} = \pp^{\tr}_{n+1}\del{\FF_n}{\vc{z}_{n}}{}. \label{eq_ppp}
\end{align}
Owing to the definition of $\am_n = \WW^{(n)}\vc{z}_{n}$, 
$\pp^{\tr}_n$ has the following relationship with 
the error to be back-propagated, $\ee^{\tr}_n = \pdel{L}{\am_n}{}$,
\begin{align}
\pp^{\tr}_{n} = \del{L}{\vc{z}_{n}}{} 
= \del{L}{ \am_n}{} \del{\am_n}{\vc{z}_n}{} = \ee_n^\tr \WW^{(n)}. 
\label{eq_adjoint_and_error}
 \end{align}
Considering Eq.~(\ref{eq_adjoint_and_error}), Eq.~(\ref{eq_ppp}) is rewritten 
as follows:
\begin{align}
\ee_n^\tr = \ee^\tr_{n+1} \WW^{(n+1)}\del{\FF_n}{\am_n}{}. \label{eq_backprop_error}
\end{align}
Hence, the gradient of $L$ with respect to the element $W_{ij}^{(n)}$ of the weight matrix $\WW^{(n)}$ is expressed as follows: 
\begin{align}
\dfrac{dL}{d W_{ij}^{(n)}} = \ee_{n+1}^\tr \WW^{(n+1)}
\dfrac{\partial \FF_n}{\partial W_{ij}^{(n)}}. \label{eq_backprop_update}
\end{align}
In the gradient descent method for training, the update of the weight matrix is expressed such that $\delta W_{ij}^{(n)} \propto -{dL}/{d W_{ij}^{(n)}}$. 
Hence, the update $\delta W_{ij}^{(n)}$ can be computed by solving Eqs.~(\ref{eq_backprop_error}) and (\ref{eq_backprop_update}). 

When considering the output in the multilayered NN, $\vc{z}_{n}$, 
as the dynamical variable ($N_r$ dimensional system state vector) 
of a discrete-time dynamical system,
the layer-to-layer information propagation in the NN is expressed as the time-evolution of the dynamical system governed by a nonlinear function
$\mathbf{F}_n$.
The time evolution of the discrete-time dynamical system can be expressed as follows:
\begin{align}
\rr_{n+1} = \FF_n(\rr_n,\vc{u}_n), \label{eq_disdys}
\end{align}
where $\rr_n \in \mathbb{R}^{N_r}$ and $\vc{u}_n \in \mathbb{R}^{N_u}$ represent the system state vector and tuning parameter vector at the $n$-th time step, respectively. The input data $\xx$ is encoded as $\rr_0 = \GG_\mathrm{in}(\xx)$, and the final state $\rr_N$ after $N$ steps is used to generate an output $\yy = \GG_\mathrm{out}(\rr_N,\ww)$, where $\GG_\mathrm{in}$ and 
$\GG_\mathrm{out}$ are functions required to convert between input/output data and the system state vector in the input and output processes, respectively.
$\ww$ denotes the weight parameters of the output function.
The loss function is defined as $L(\yy,\tm)$.
By minimizing $L(\yy,\tm)$ through optimization of the parameters $\vc{u}_n$ and $\ww$, the dynamical system performs as effectively as the multilayered NN.

If $\pp_{n}$ is defined as
\begin{align}
\pp_{n}^\tr = \del{L}{\rr_{n}}{}, \label{eq_pp_dLdr}
\end{align}
the gradient vector for the optimization is expressed as,
\begin{align}
\dev{L}{\vc{u}_n}{} 
&= 
\del{L}{\rr_{n+1}}{} \dev{\rr_{n+1}}{\vc{u}_n}{} 
= \del{L}{\rr_{n+1}}{} \del{\FF_{n}}{\vc{u}_n}{} \nonumber \\
&= \pp_{n+1}^\tr \del{\FF_{n}}{\vc{u}_n}{}. \label{eq_update_backprop_p}
\end{align}
%
The vector $\pp_{n}$ can be computed backward from the final state $\pp_{N}$
using the following equation recurrently:
\begin{align}
\pp_{n}^{\tr}
&= \del{L}{\rr_{n}}{} = \del{L}{\rr_{n+1}}{}\del{\FF_n}{\rr_{n}}{} \nonumber \\
&= \pp_{n+1}^\tr\del{\FF_n}{\rr_{n}}{}. \label{eq_backprop_p}
\end{align}
Thus, 
the update $\delta \vc{u}^{\tr}_n \propto -\pdev{L}{\vc{u}_n}{}$ can be computed by solving Eqs.~(\ref{eq_update_backprop_p}) and (\ref{eq_backprop_p}). 
As shown in Sec.~\ref{sec_2}, $\pp_n$ corresponds to {\it the adjoint vector}, and Eq.~(\ref{eq_backprop_p}) corresponds to {\it the adjoint equation}, which can be used for the optimal control of a discrete-time dynamical system.

\subsection{Direct feedback alignment (DFA)}
In the DFA method for multilayered NNs, the error vector $\ee_n$ and weight matrix $\WW^{(n+1)}$ are replaced with the error at the last layer $\ee = \pdel{L}{\am_N}{} \in \mathbb{R}^{L}$ and a random matrix $\BB_{n} \in \mathbb{R}^{L \times N_r}$, respectively.
Considering Eq.~(\ref{eq_adjoint_and_error}), $\pp_n$ is replaced by 
\begin{align}
\hat{\pp}^{\tr}_{n} = \ee^\tr \BB_n.
\end{align}
The error $\ee$ is given by $\yy - \tm$ when the loss function $L$ is the MSE for a regression task as $L = 1/2(\yy-\yy_\mathrm{tag})^2$ or the cross entropy for a classification task as $L = -\tm \log\yy$.

With this expression, the matrix element $\hat{W}_{ij}^{(n)}$ is updated as follows:
\begin{align}
\delta \hat{W}_{ij}^{(n)} 
&\propto -\ee^\tr \BB_{n+1}
\dfrac{\partial \FF_n}{\partial W_{ij}^{(n)}}.
\label{eq_dfa_st_update}
\end{align}

When applying this scheme for multilayered NNs 
directly into the dynamical system, 
the vector $\pp_n^{\tr}$ outlined in Eq.~(\ref{eq_pp_dLdr}) can 
be replaced by
\begin{align}
\hat{\pp}_n^{\tr} = \ee^\tr \BB_n.  
\end{align}
Hence, $\hat{\uu}^{(n)}$ is updated as follows:
\begin{align}
\delta \hat{\uu}_n^{\tr} \propto 
-\hat{\pp}_{n+1}^{\tr} \del{\FF_n}{\uu_n}{}
= -\ee^\tr \BB_{n+1} \del{\FF_n}{\uu_n}{}.
\label{eq_dfa_general_update}
\end{align}

\section{Adjoint equations for discrete time dynamical systems \label{sec_2}}
Here, we apply the adjoint method developed for optimal control of dynamical systems to Eq.~(\ref{eq_disdys}) and obtain the adjoint equations.
The derived adjoint equation and gradient vector correspond to Eq.~(\ref{eq_backprop_p}) for backpropagation and Eq.~(\ref{eq_update_backprop_p}), respectively.
Derivation proceeds by first incorporating the constraint of Eq.~(\ref{eq_disdys}) into 
an augmented loss function $L_\mathrm{aug}$ with the Lagrangian multiplier vector, $\pp_n \in \RR^{N_r}$. 
The augmented loss function is given as 
\begin{align}
L_\mathrm{aug} = 
L(\yy, \tm)
+ \sum_{n=0}^{N-1}
\pp^\tr_{n+1}
\left(\FF_n - \rr_{n+1}
\right). \label{eq_adjoint_dis}
\end{align}
Notably, $L_\mathrm{aug} = L(\yy, \tm)$ for all valid trajectories of $\rr_n$ obeying Eq.~(\ref{eq_disdys}). The second term of Eq.~(\ref{eq_adjoint_dis}) can be rewritten as follows:
\begin{align}
\sum_{n=0}^{N-1}
\pp^\tr_{n+1}
\left(\FF_n - \rr_{n+1}
\right)
= 
\sum_{n=0}^{N-1}
\pp^\tr_{n+1}\FF_{n} 
-
\sum_{n=1}^{N}
\pp^\tr_{n}\rr_{n} 
=
\sum_{n=1}^{N-1}
\left(
\pp^\tr_{n+1}\FF_{n} 
-
\pp^\tr_{n}\rr_{n}
\right)
+\pp^{\tr}_{1}\rr_0-\pp^{\tr}_N\rr_N,
\end{align}
and we have, 
\begin{align}
L_\mathrm{aug} 
&= 
L
+
\sum_{n=1}^{N-1}
\left(
\pp^\tr_{n+1}\FF_{n} 
-
\pp^\tr_{n}\rr_{n}
\right)
+\pp^{\tr}_{1}\rr_0-\pp^{\tr}_N\rr_N.
\label{eq_adjoint_dis_2}
\end{align}
The gradient vector is computed as follows:
\begin{align}
\dev{L}{\uu_{n}}{} 
=
\dev{L_\mathrm{aug}}{\uu_{n}}{} 
&=
\del{L}{\rr_N}{}
\dev{\rr_N}{\uu_{n}}{}
+
\sum_{n'=1}^{N-1}
\pp^\tr_{n'+1}
\left(
\del{\FF_{n'}}{\rr_{n'}}{}\dev{\rr_{n'}}{\uu_{n}}{} 
+ 
\del{\FF_{n'}}{\uu_{n}}{}
\right)
 -
\sum_{n'=1}^{N-1}
\pp^\tr_{n'}
\dev{\rr_{n'}}{\uu_{n}}{} 
-
\pp^{\tr}_N\dev{\rr_{N}}{\uu_{n}}{} 
\nonumber \\
&=
\left(
\del{L}{\rr_N}{}
- \pp_N^\tr
\right)
\dev{\rr_N}{\uu_{n}}{}
+
\sum_{n'=1}^{N-1}
\left(
\pp_{n'+1}^\tr
\del{\FF_{n'}}{\rr_{n'}}{}
- 
\pp_{n'}^\tr
\right)\dev{\rr_{n'}}{\uu_{n}}{} 
+
\sum_{n'=1}^{N-1}
\pp^\tr_{n'+1}
\del{\FF_{n'}}{\uu_{n}}{} \nonumber \\
&=
\left(
\del{L}{\rr_N}{}
- \pp_N^\tr
\right)
\dev{\rr_N}{\uu_{n}}{}
+
\sum_{n'=1}^{N-1}
\left(
\pp_{n'+1}^\tr
\del{\FF_{n'}}{\rr_{n'}}{}
- 
\pp_{n'}^\tr
\right)\dev{\rr_{n'}}{\uu_{n}}{} 
+
\pp^\tr_{n+1}
\del{\FF_{n}}{\uu_{n}}{},
\end{align}
where $\pdev{\rr_0}{\uu_{n}}{} = 0$ is utilized because the initial state $\rr_0$ is independent on $\uu_{n}$. 
In addition, the third term is derived from $\pdel{\FF_{n'}}{\uu_{n}}{} = \pdel{\FF_{n}}{\uu_{n}}{}\delta_{n'n}{}$, where $\delta_{n'n}$ denotes the Kronecker's delta. 
 
The Lagrangian multiplier $\pp_n$ can be set freely; therefore, we can select $\pp_n$ such that it satisfies the following equations:
\begin{align}
\pp^\tr_{N} = \del{L}{\rr_N}{}, \label{eq_adjoint_disdyn_N}
\end{align}
and 
\begin{align}
\pp^\tr_{n}
=
\pp^\tr_{n+1}\del{\FF_n}{\rr_n}{}, 
\label{eq_adjoint_disdyn}
\end{align}
for $n \in \{1, \cdots, N-1\}$. 
This results in a simple form of $\pdev{L_\mathrm{aug}}{\uu_{n}}{}$ as,
\begin{align}
\dev{L_\mathrm{aug}}{\uu_{n}}{} 
=
\pp^\tr_{n+1}\del{\FF_{n}}{\uu_{n}}{}.
\end{align}
Because $\pdev{L}{\uu_n}{} = \pdev{L_\mathrm{aug}}{\uu_n}{}$ in the gradient descent scheme, the update $\delta \uu_n$ is given as, 
\begin{align}
\delta \uu^{\tr}_n \propto -\pp^\tr_{n+1}\del{\FF_{n}}{\uu_{n}}{}.
\label{eq_dfa_update_discrete}
\end{align}
In summary, when we set the adjoint vector $\pp_n$ such that it satisfies Eqs.~(\ref{eq_adjoint_disdyn_N}) and (\ref{eq_adjoint_disdyn}), 
the adjoint Eq.~(\ref{eq_adjoint_disdyn}) corresponds to backward Eq.~(\ref{eq_backprop_p}), suggesting $\pp^{\tr}_n = \pdel{L}{\rr_n}{}$; thus, we can obtain the same update equation as Eq.~(\ref{eq_update_backprop_p}). 

\subsection{DFA for adjoint method in discrete-time dynamical systems}
In this subsection, we describe the DFA-based training for the adjoint method, based on the correspondence between backpropagation and the adjoint method.
Concretely, the adjoint vector $\pp_n$ is replaced by $\hat{\pp}^{\tr}_n = \ee^\tr \BB_n$, where $\ee = \yy - \tm$ and $\BB_n$ denote the error vector in the last layer and a random matrix, respectively. 
By substituting this into Eq.(\ref{eq_dfa_update_discrete}), the update rule based on the DFA scheme is given as follows:
\begin{align}
\delta \hat{\uu}_n^{\tr} 
&\propto -\hat{\pp}_{n+1}^{\tr}
\del{\FF_n}{\uu_n}{} \nonumber \\
&= -\ee^\tr\BB_{n+1}\del{\FF_n}{\uu_n}{}. 
\end{align}
We call this training method and the vector $\hat{\pp}$ as {\it DFA--adjoint} method and DFA--adjoint vector, respectively.

\section{Adjoint equations for continuous-time dynamical systems}
We consider a continuous-time dynamical system.
It is assumed that the system state $\rr(t) \in \mathbb{R}^{N_r}$ can be controlled by continuously altering the control vector $\uu(t) \in \mathbb{R}^{N_u}$ and is governed by the following equation:
\begin{align}
\dev{\rr(t)}{t}{} = \FF[\rr(t),\uu(t)]. \label{eq_ctdyn}
\end{align}
The system state $\rr(t)$ evolves starting from an initial state synthesized from input $\xx$ as $\rr(0) = \GG_\mathrm{in}(\xx)$.
The input information is transformed through the time evolution up to time $t = T_e$. 
The output vector $\yy \in \mathbb{R}^L$ is given 
with the final state $\rr(T_e)$ as $\yy = \GG_\mathrm{out}[\rr(T_e)]$. 
The learning process aims at optimizing the control signal $\uu(t)$ such that the loss function $L(\yy,\tm)$ is minimized for a given training dataset. 

The augmented loss function $L_\mathrm{aug}$ with the Lagrangian multiplier (adjoint) vector $\pp(t)$ is given as 
\begin{align}
L_\mathrm{aug} = 
L(\yy, \tm)
+ \int_0^{T_e}
\pp^\tr(t)
\left\{ \FF[\rr(t),\uu(t)] - \dev{\rr(t)}{t}{}
\right\} dt. \label{eq_adjoint_cs}
\end{align}
The second term of the aforementioned equation can be rewritten as 
\begin{align}
\int_0^{T_e}
\pp^\tr(t)
\left(
\FF(\rr,\uu)
-\dev{\rr(t)}{t}{}
\right)dt
= 
-\left[\pp^{\tr}(t)\rr(t)\right]_{0}^{T_e}
+ 
\int_0^{T_e}
\left(
\pp^{\tr}(t)\FF(\rr,\uu) + 
\dev{{\pp}^{\tr}(t)}{t}{}\rr(t)
\right)
dt,
\end{align}
and the augmented loss function $L_\mathrm{aug}$ is as follows:
\begin{align}
L_\mathrm{aug} = 
L(\yy, \tm)
-\pp^{\tr}(T_e)\rr(T_e)
+
\pp^{\tr}(0)\rr(0)
+ 
\int_0^{T_e}
\left(
\pp^{\tr}(t)\FF(\rr,\uu) + 
\dev{{\pp}^{\tr}(t)}{t}{}\rr(t)
\right)
dt.
\label{eq_adjoint_cs_ver2}
\end{align}
In terms of variations $\delta L$ and $\delta \rr(t)$ with respect to $\delta \uu(t)$, we can obtain,
\begin{align}
\delta L_\mathrm{aug}
&=
\left\{
\del{L}{\rr(T_e)}{}
- \pp^\tr(T_e)
\right\}
\delta{\rr(T_e)}
+
\int_0^{T_e}
\left\{
\pp^\tr(t)
\del{\FF}{\rr}{}
+ 
\dev{\pp^\tr}{t}{}
\right\} \delta \rr(t) dt
+
\int_0^{T_e}
\pp^\tr(t)
\del{\FF}{\uu}{}\delta\uu(t) dt, 
\end{align}
where $\delta \rr(0) = 0$ is used because we assume that $\rr(0) = \GG_\mathrm{in}(\xx)$ is fixed in this case. 
When we choose $\pp(t)$ such that $\pp^\tr(T_e) = \pdel{L}{\rr(T_e)}{}$ and the following equation is satisfied, 
\begin{align}
\dev{\pp^\tr(t)}{t}{}
= 
-\pp^\tr(t)
\del{\FF}{\rr}{}, 
\end{align}
$L_\mathrm{aug}$ is simplified as $\int_0^{T_e}\pp^\tr(t)\pdel{\FF}{\uu}{}\delta\uu(t)dt$.
Accordingly, $\delta L_\mathrm{aug} \le 0$ when we can choose $\delta\uu(t)$ as 
\begin{align}
\delta\uu^{\tr}(t) \propto 
- 
\pp^\tr(t)
\del{\FF}{\uu}{}.
\end{align}

\subsection{DFA--adjoint for continuous-time dynamical systems}
Similar to the DFA--adjoint in discrete-time systems, the connection with the DFA method in continuous-time systems is made by replacing the adjoint vector $\pp^\tr(t)$ with the DFA--adjoint vector $\hat{\pp}^{\tr}(t)$. This vector is defined as 
\begin{align}
\hat{\pp}^{\tr}(t) = \ee^T\BB(t),
\end{align}
 where $\ee = \yy - \tm \in \mathbb{R}^L$ and $\BB(t) \in \mathbb{R}^{L \times N_r}$ denote the error vector in the last layer and a time-dependent matrix, respectively. 
We assume that the time variation of $\BB(t)$ is not limited to random in continuous-time systems, unlike the conventional DFA method.   
In Sec.~\ref{sec_bb}, we set $\BB(t)$ as a various time-dependent function and evaluated the performance. 
Because the control signal $\uu(t)$ is dependent on $\BB(t)$ via $\hat{\pp}(t)$, the characteristic timescale of $\BB(t)$ should be close to that of the dynamical system. 

The DFA--adjoint method can be applied to arbitrary continuous-time dynamical systems. 
Given the equivalence between multilayered NNs and dynamical systems, utilizing high-dimensional dynamical systems is crucial for achieving optimal performance. However, finding high-dimensional dynamical systems with numerous controllable parameters for physical implementations poses a significant challenge. To address this issue, we introduce a dynamical system with a time delay in the subsequent section.
 
\section{Adjoint equations for continuous-time systems with time delay}
In dynamical systems with time delay, the degrees of freedom in the time domain can be exploited as virtual nodes, which can be leveraged for computational purposes. Consequently, physical systems with time delay are useful for facilitating large-scale computations even with a simple physical configuration. In this section, we delve into the dynamical systems with time delay and derive the adjoint equation [Eq.~(2) in the main text] and the update rule for the control signal $\uu(t)$. Additionally, we also derive the update rule for the static system parameter $\ttheta$ and input mask $\mm(t)$ for a more general formulation. 

To this end, we begin with the following $N_r$ dimensional equation, 
\begin{align}
\dev{\rr(t)}{t}{} = \FF \left[ \rr(t),\rr(t-\tau_D),\uu(t),\ttheta \right].
\end{align}
We assumed that the input $\xx \in \mathbb{R}^{N_x}$ is encoded 
by a matrix $\mm(t) \in \mathbb{R}^{N_x \times N_r}$ as follows:
\begin{align}
\rr(t) = \GG_\mathrm{in}(\xx,t) = \mm^\tr(t)\xx, \mbox{ for $-\tau_D \le t \le 0$},
\label{eq_mask}
\end{align}
and the output $\yy \in \mathbb{R}^L$ is given by 
\begin{align}
\yy = \GG_\mathrm{out}(\zz), \\
\zz = \int_{T_e-T_c}^{T_e} \ww(t) \rr(t) dt,
\end{align}
where $\ww(t) \in \mathbb{R}^{L\times N_r}$ represents an weight matrix to produce the output, and
$T_c$ represents the time length to extract the output from the time series.

The augmented loss function is given by 
\begin{align}
L_\mathrm{aug} = 
L\left[
\tm,\yy\left(\zz\right)
\right]
+
\int_0^{T_e}
\pp^\tr(t)
\left\{
\FF[\rr,\rr(t-\tau_D),\uu,\ttheta]
-\dev{\rr(t)}{t}{}
\right\}dt.
\end{align}
Because the second term of the aforementioned equation can be rewritten as 
\begin{align}
\int_0^{T_e}
\pp^\tr(t)
\left(
\FF
-\dev{\rr(t)}{t}{}
\right)dt
= 
-\left[\pp^{\tr}(t)\rr(t)\right]_{0}^{T_e}
+ 
\int_0^{T_e}
\left(
\pp^{\tr}(t)\FF + 
\dev{{\pp^{\tr}}(t)}{t}{}\rr(t)
\right)
dt,
\end{align}
the variations $\delta L_\mathrm{aug}$ and $\delta \rr(t)$ in terms of $\delta\uu(t)$, $\delta\ttheta$, and $\delta\mm(t)$ can be derived as follows:
\begin{align}
  \dl L_\mathrm{aug} &=
\del{L}{\zz}{}\dl\zz
  -
\left\{
\pp^{\tr}(0)\dl \rr(0)
  -\pp^{\tr}(T_e)\dl \rr(\Te)
\right\}
+
    \int_{0}^{T_e}
    \left(
\dev{{\pp}^{\tr}}{t}{} +  \pp^{\tr}\del{\FF}{\rr}{}
    \right)\dl \rr dt \nonumber \\
  &+
    \int_{0}^{T_e}
    \pp^{\tr}(t)\del{\FF}{\rr_{\tau_D}}{}\dl \rr_{\ta_D}
   dt
  +
  \int_0^{T_e}\pp^{\tr}\del{\FF}{\uu}{}\dl \uu dt 
+
\int_0^{T_e}\pp^{\tr}\del{\FF}{\ttheta}{}\dl\ttheta dt, \label{ad_eq1}
\end{align}
where
$\rr_{\tau_D} = \rr(t-\tau_D)$ and 
\begin{align}
\dl\zz = \int_{T_e-T_c}^{T_e}\ww(t)\dl\rr dt 
= \int_{0}^{T_e}\Theta_1(t)\ww(t)\dl\rr dt.
\end{align}
$\Theta_1(t)$ represents the indicator function, which equals 1 for $T_e-T_c \le t \le T_e$ and 0 otherwise. 
The third term in Eq.~(\ref{ad_eq1}) can be rewritten as follows: 
\begin{align}
 \int_{0}^{T_e}
    \pp^{\tr}\del{\FF}{\rr_{\tau_D}}{}\dl \rr_{\ta_D}
   dt
&= 
    \int_{-\tau_D}^{T_e-\tau_D}
    \pp^{\tr}(t+\tau_D)\del{\FF(t+\tau_D)}{\rr}{}\dl\rr
   dt \nonumber \\
&=
    \int_{0}^{T_e}
    \Theta_0(t)\pp^{\tr}(t+\tau_D)\del{\FF(t+\tau_D)}{\rr}{}\dl\rr
   dt
+
    \int_{-\tau_D}^{0}
    \pp^{\tr}(t+\tau_D)\del{\FF(t+\tau_D)}{\rr}{}\dl\rr
   dt,
\end{align}
where $\Theta_0(t)$ represents the indicator function, which equals 1 for $0 < t \le T_e-\tau_D$ and 0 otherwise.
%
Therefore, the augmented loss function is rewritten as
\begin{align}
  \dl L_\mathrm{aug} =
&-\pp^{\tr}(T_e)\dl \rr(\Te) \nonumber \\
&+
\int_{0}^{T_e}
\left[
\Theta_1(t)\del{L}{\zz}{}
\ww
+
\dev{{\pp}^{\tr}}{t}{} +  \pp^{\tr}\del{\FF}{\rr}{}
+
\Theta_0(t)\pp^{\tr}(t+\tau_D)\del{\FF(t+\tau_D)}{\rr}{}
\right]\dl\rr dt \nonumber \\
&+
\int_0^{T_e}\pp^{\tr}\del{\FF}{\uu}{}\dl\uu dt 
+
\int_0^{T_e}\pp^{\tr}\del{\FF}{\ttheta}{}\dl\ttheta dt 
+
\int_{-\tau_D}^{0}
\pp^{\tr}(t+\tau_D)\del{\FF(t+\tau_D)}{\rr}{}\dl\mm^{\tr}\xx dt. 
\end{align}
We can choose the adjoint vector $\pp(t)$ such that the following equation is satisfied:
\begin{align}
\dev{\pp^{\tr}}{t}{}
= 
-\Theta_1(t)\del{L}{\zz}{}
\ww
-\pp^{\tr}\del{\FF}{\rr}{}
-
\Theta_0(t)\pp^{\tr}(t+\tau_D)\del{\FF(t+\tau_D)}{\rr}{}.
\label{eq_adjoint_delay_1}
\end{align}
In this case, the variation in the loss function is
\begin{align}
  \dl L_\mathrm{aug} =
\int_0^{T_e}\pp^{\tr}\del{\FF}{\uu}{}\dl\uu dt 
+
\left(
\int_0^{T_e}\pp^{\tr}\del{\FF}{\ttheta}{}dt
\right)\dl\ttheta
+
\int_{-\tau_D}^{0}
\pp^{\tr}(t+\tau_D)\del{\FF(t+\tau_D)}{\rr}{}\dl\mm^{\tr}\xx dt. 
\end{align}
Accordingly, we can obtain $\dl L_\mathrm{aug} \le 0$ if $\uu(t)$, $\ttheta$, and $\mm(t)$ are updated as follows:
\begin{align}
\dl\uu(t) 
&= -\alpha_u 
\left(\pp^{\tr}\del{\FF}{\uu}{} 
\right)^{\tr} 
\label{eq:update-u-adjoint-timedelay}
\\
\dl\ttheta
&= 
-\alpha_{\theta}
\int_0^{T_e}
\left(
\pp^{\tr}\del{\FF}{\ttheta}{}
\right)^{\tr}
dt \\
\dl\mm^{\tr}(t)
&= -\alpha_m 
\left[
\pp^{\tr}(t+\tau_D)\del{\FF(t+\tau_D)}{\rr}{}
\right]^{\tr}
\xx^{\tr},
\end{align}
where $\alpha_u$, $\alpha_{\theta}$, and $\alpha_m$ are positive constant values.
$\uu(t)$, $\ttheta$, and $\mm^\tr(t)$ can be updated based on these update rules by solving the adjoint Eq.~(\ref{eq_adjoint_delay_1}). Instead of solving the adjoint equation, we can set 
\begin{align}
 \hat{\pp}^{\tr}(t) = \ee^{\tr}\BB(t).
\label{eq:DFA-p-timedelay}
\end{align}
The update rules based on the DFA--adjoint method are expressed as follows:
\begin{align}
\dl\uu(t) 
&= -\alpha_u
\left(
\BB(t)
\del{\FF}{\uu}{} 
\right)^{\tr} 
\ee
\label{eq:update-u-timedelay}
\\
\dl\ttheta
&= 
-\alpha_{\theta}
\int_0^{T_e}
\left(
\BB(t)
\del{\FF}{\ttheta}{} 
\right)^\tr
\ee
dt 
\\
\dl\mm^{\tr}(t)
&= 
-\alpha_m 
\left[
\BB(t+\tau_D)
\del{\FF(t+\tau_D)}{\rr}{}
\right]^{\tr}
\ee
\xx^{\tr}. \label{eq:update-dfa-delay-mask} 
\end{align}

\section{Time-dependent random matrix $\BB(t)$ \label{sec_bb}}
\begin{figure}[htbp]
\centering\includegraphics[width=8.8cm]{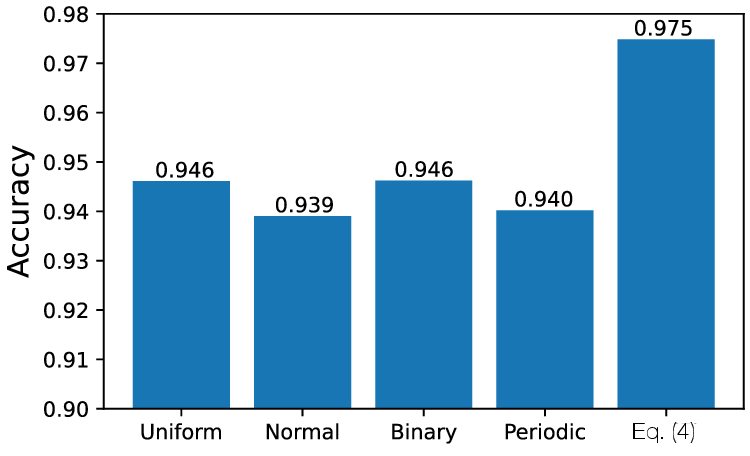}
\caption{\label{fig_s1}
Classification accuracy for the MNIST handwritten digit dataset. The time series of $\BB(t)$ was set as uniformly distributed random numbers (Uniform) within $\left[0,1\right]$, normally distributed random numbers (Normal), binary random numbers (Binary), a periodic waveform with a period of 85.03 $\mu$s (Periodic), and the waveform shown in Eq.~(4) in the main text.
}
\end{figure}

In this section, we set various time-dependent random matrices $\BB(t)$ and evaluated the classification accuracy for the MNIST handwritten digit image dataset under different scenarios [Fig.~\ref{fig_s1}]. 
In three distinct cases, the elements $B_{li}(t)$ of the time-dependent random matrix were generated using uniformly distributed random numbers within $\left[0,1\right]$, normally distributed random numbers with zero mean and unit variance, and binary random numbers $\{-1,1\}$, such that $B_{li}$ satisfies $\langle (B_{li}(t) - \bar{B})(B_{l'i'}(t') - \bar{B})\rangle_t = \sigma^2\delta(t-t')\delta_{ll'}\delta_{ii'}$, where $\bar{B}$ and $\sigma^2$ represent the mean and variance, respectively. 
$l$ and $i$ denote the indices of a target class and the degree of freedom of $\rr(t)$, respectively. 
In addition, we introduced a periodic waveform defined as $B_{li}(t) = B^0_{li} \sin(\omega t + 2\pi l/10)$ and explored the performance for various angular frequencies. 
Notably, we determined that, when $2\pi/\omega = 85.03$ $\mu$s is close to the timescale $\tau_L$, the classification accuracy reached 0.94.

We used the following $\BB(t)$ in the main text, 
\begin{align}
B_{li}  = \sum_{m=1}^{M_B}A_m^{(li)}\sin\left(
\omega_m t + 2\pi\phi_{m}^{(li)}
\right), \label{eq_b_matrix}
\end{align}
and we obtained an accuracy of more than 0.97, surpassing the performance of the other configurations.
Here, $A_m^{(li)}$ and $\phi_m^{(li)}$ are random numbers sampled from the continuous uniform distribution within $[0,1)$. The angular frequencies were determined by $\omega_m = 2\pi(14m/M_B + 1)$, where the unit for the angular frequency is radian per millisecond. $M_B = 3\times 28^2$ was used. 

\begin{figure}[htbp]
\centering\includegraphics[width=8.8cm]{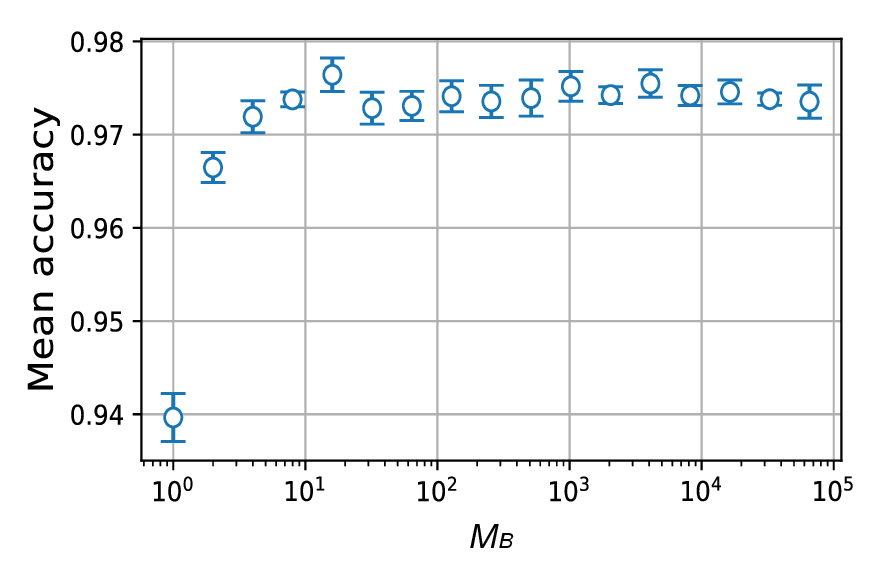}
\caption{\label{fig_s2}
Classification accuracy as a function of $M_B$. 
}
\end{figure}

Figure~\ref{fig_s2} shows the classification accuracy as a function of $M_B$ in Eq.~(\ref{eq_b_matrix}). For $M_B \ge 4$, the accuracy was saturated to be approximately 97$\%$.

\section{Toward a model-free training approach}
The drawback of the DFA--adjoint method is the requirement of detailed prior knowledge of dynamical operator $\FF$ in the physical system. For example, $\pdel{\FF}{\uu}{}$ is used for the update of control $\uu$, as shown in Eq.~(\ref{eq:update-u-timedelay}). This may make it difficult to use the DFA--adjoint method for training the system. To address this issue, we introduce two different approaches in this section.  
\subsection{Augmented DFA--adjoint method}
We assume that we do not possess an accurate model of the physical system, $\dot{\rr} = \FF(\rr,\rr(t-\tau_D),\uu)$; instead, we have an approximate model, represented as $\dot{\rr} \approx \HH(\rr,\rr(t-\tau_D),\uu)$. In this case, we can use $\HH(\rr,\rr(t-\tau_D),\uu)$ to compute the derivatives required for the update in Eqs.~(\ref{eq:update-u-timedelay})--(\ref{eq:update-dfa-delay-mask}) instead of using $\FF(\rr,\rr(t-\tau_D),\uu)$.
In this method, $\uu$ is updated as follows:
\begin{align}
\uu(t) \rightarrow \uu(t) - \alpha_u\dl\uu_{aug}(t),
\end{align}
where
\begin{align}
\dl\uu_{aug}(t) 
&= -\alpha_u
\left(
\BB(t)
\del{\HH}{\uu}{} 
\right)^{\tr} 
\ee.
\label{eq:update-adfa-u-timedelay}
\end{align}
If the physical system depends on parameter $\ttheta$, we use the model $\HH(\rr,\rr(t-\tau_D),\uu,\ttheta)$. When the parameter $\ttheta$ and mask pattern $\mm(t)$ are also trained, the updates $\dl\ttheta_{aug}$ and $\dl\mm_{aug}(t)$ are expressed as follows:
\begin{align}
\dl\ttheta_{aug}
&= 
-\alpha_{\theta}
\int_0^{T_e}
\left(
\BB(t)
\del{\HH}{\ttheta}{} 
\right)^\tr
\ee
dt 
\\
\dl\mm^{\tr}_{aug}(t)
&= 
-\alpha_m 
\left[
\BB(t+\tau_D)
\del{\HH(t+\tau_D)}{\rr}{}
\right]^{\tr}
\ee
\xx^{\tr}. \label{eq:update-adfa-delay-mask} 
\end{align}
This method is inspired by the so-called augmented DFA [10], where the derivatives of the activation function in a neural network are replaced with an appropriate function that correlates with the original activation function. In this sense, we refer to the proposed method as the augmented DFA--adjoint method in this study. The augmented DFA--adjoint method is more general than the augmented DFA because it can be applied to a wide range of physical systems where neurons and synaptic weights cannot be explicitly defined.

Table~\ref{tab:table_augmentedDFA} presents the classification accuracy achieved with the augmented DFA--adjoint method, using the Gaussian, $\tanh$, and Softplus (a smooth approximation of ReLU) functions as $\HH$. These functions are typically used as activation functions in conventional neural networks but differ from the original $\FF$. 
The factors, $\am_0$, $\am_1$, and $\am_2$, are used to compensate for the timescale of the system and learning rate. 
We found that the augmented DFA--adjoint method with the Gaussian and $\tanh$ models achieves classification accuracy close to that of the DFA--adjoint method. In the case of the Softplus function model ($\log(1+e^\delta)$), the accuracy degraded to approximately 91$\%$, but this is still better than that achieved with the ELM. 
These results suggest that a precise model of the physical system is not necessary; an approximate model of $\FF$ is sufficient for training. This may be attributed to the error robustness of the proposed approach, as discussed in the main text.

\begin{table}[b]
\caption{\label{tab:table_augmentedDFA}
Comparison of classification accuracy between the DFA--adjoint and augmented DFA--adjoint methods using $\HH = \pm\am_0\exp(-\delta^2)$, $\am_1\tanh(\delta + \pi/4)$, and $\am_2\log(1+e^\delta)$. $\delta = u_1(t)r_1(t-\tau_D) + u_2(t)$, $\am_0 = (1/\tau_L,1/\tau_L)^\tr$, $\am_1 = (1/\tau_L,1/\tau_L)^\tr/10$, $\am_2 = (1/\tau_L,1/\tau_L)/10^{4}$. The MNIST dataset was used for this comparison.}
\begin{ruledtabular}
\begin{tabular}{lcdr}
\textrm{Methods }&
Model&
\multicolumn{1}{c}{\textrm{Accuracy}}\\
\colrule
DFA--adjoint & Eqs.~(\ref{eq_Murphy_1}) and (\ref{eq_Murphy_2}) &0.973\\
Augmented DFA--adjoint & $\am_0\exp(-\delta^2)$ &0.970\\
Augmented DFA--adjoint & $-\am_0\exp(-\delta^2)$ &0.970\\
Augmented DFA--adjoint & $\am_1\tanh(\delta + \pi/4)$ &0.967\\
Augmented DFA--adjoint & $\am_2\log(1+e^\delta)$ &0.906\\
ELM (w/o control)  & -& 0.869
\end{tabular}
\end{ruledtabular}
\end{table}

\subsection{Additive control-based DFA--adjoint method}
The proposed DFA--adjoint method is based on an external control approach. Thus, we can flexibly design how to apply control signal $\uu(t)$ to the physical system, so that $\uu$ can be trained without computing the derivative of $\FF$ on $\uu$. A straightforward case is that control signal $\uu$ is additively applied to a physical system. We assume that the model of the physical systems without control, $\dot{\rr} = \FF_0(\rr,\rr(t-\tau_D))$, is unknown. 
For the additive control, the equation is given as follows:
\begin{align}
\dev{\rr}{t}{} = \FF\left[
\rr,\rr(t-\tau_D),\uu
\right]
= \FF_0\left[
\rr,\rr(t-\tau_D)
\right]
 + \alpha \uu(t), 
\end{align}
where $\alpha$ is a constant. In this case, we explicitly compute $\pdel{\FF}{\uu}{}$ as follows:
\begin{align}
\del{\FF}{\uu}{} = \alpha\II,
\end{align}
where $\II$ denotes an identity matrix. Thus, the update of $\uu(t)$ is given by
\begin{align}
 \uu(t) \rightarrow \uu(t) - \eta_u\left(
\del{\FF}{\uu}{}
\right)^\tr
\pp(t) = \uu(t) - \tilde{\eta}_u\pp(t)
\end{align}
where $\tilde{\eta}_u = \eta_u\alpha$ is used as the learning rate. 
For the DFA--adjoint method, the update is given by 
\begin{align}
 \uu(t) \rightarrow \uu(t) - \tilde{\eta}_u\BB^T(t)\ee.
\end{align} 

\begin{figure}[htbp]
\centering\includegraphics[width=8.8cm]{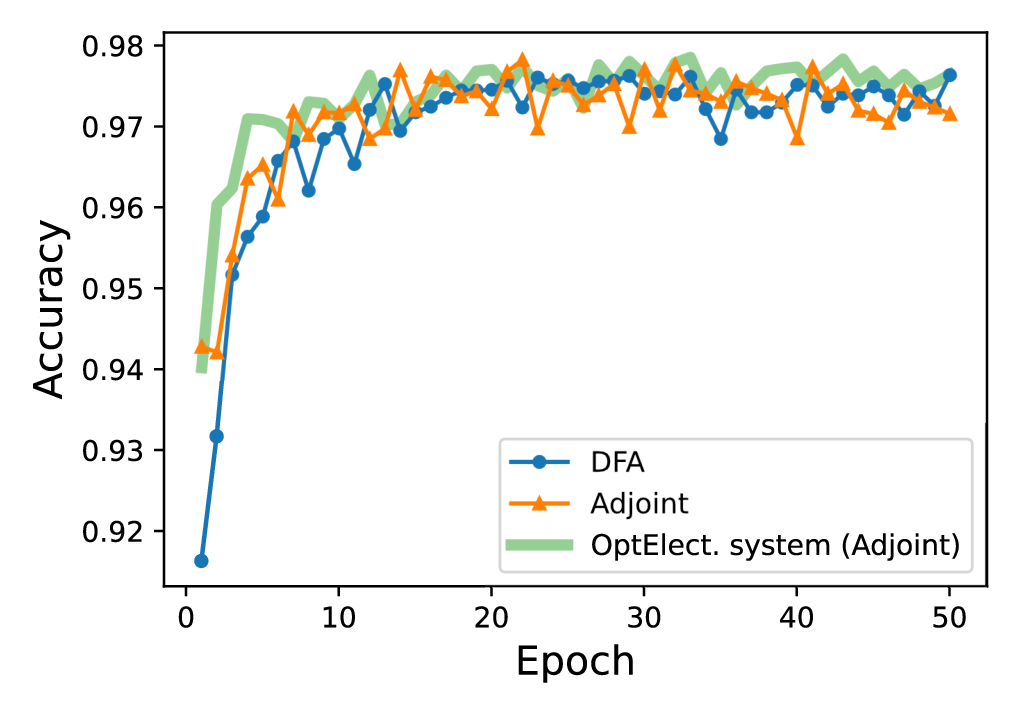}
\caption{\label{fig_s3}
Classification accuracy for the MNIST handwritten digit dataset. 
Blue circles: DFA--adjoint method, orange triangles: adjoint method applied to additive control model [Eq.~(\ref{eq:delay-additive-model})]. Green line: the adjoint method applied to Eq.~(5) in the main text. $\alpha$ = 1.}
\end{figure}

To investigate the effectiveness of the additive-control-based DFA--adjoint method, we used the following optoelectronic system model with additive control $u_1(t)$,
\begin{align}
\tau_L \dev{r}{t}{} 
= -\left(1 + \dfrac{\tau_L}{\tau_H} \right)r
- \dfrac{1}{\tau_H}\int rdt
+ \beta\cos^2\left[ r(t - \tau_D) - \pi/4 \right] + \alpha u_1(t). 
\label{eq:delay-additive-model}
\end{align}
Figure~\ref{fig_s3} shows the classification accuracy as a function of epoch. We used the adjoint and DFA--adjoint methods for the additive-control model.
For comparison, we also used the original optoelectronic system model [Eq.~(5) in the main text] trained by the adjoint method. 
The classification accuracy was 0.9723 for the adjoint method and 0.9737 for the DFA--adjoint method. The accuracies were comparable with that for the original model (Fig.~\ref{fig_s3}).

\section{Numerical simulation}

The model equation of the optoelectronic delay system utilized for 
the numerical simulations in this study is represented as Eq.~(5) in the main text.
The original model equation proposed by Murphy et al. is given as follows:
\begin{align}
\dev{r_1}{t}{} 
&= -\left( \frac{1}{\tau_L} + \frac{1}{\tau_H} \right) r_1(t) - r_2(t)
+ \frac{\beta}{\tau_L} \cos^2\left[ r_1(t-\tau_D) + \phi_0 \right],
\\
\dev{r_2}{t}{} &= \frac{1}{\tau_H} r_1(t).
\label{eq_Murphy}
\end{align}
To apply the system for the present training scheme,
two control signals $u_1(t)$ and $u_2(t)$ were introduced
to impact $r_1(t - \tau_D)$ and $\phi_0$, respectively, as follows:
\begin{align}
\dev{r_1}{t}{} 
&= -\left( \frac{1}{\tau_L} + \frac{1}{\tau_H} \right) r_1(t) - r_2(t)
+ \frac{\beta}{\tau_L} \cos^2\left[ u_1(t)r_1(t-\tau_D) + u_2(t) \right],
\label{eq_Murphy_1}
\\
\dev{r_2}{t}{} &= \frac{1}{\tau_H} r_1(t).
\label{eq_Murphy_2}
\end{align}
Additionally, a noise term, $\sigma \xi(t)$, was incorporated into
the system to explore noise robustness,
where $\sigma$ and $\xi(t)$ represent the noise strength and Gaussian noise
with zero mean and unity standard deviation, respectively.
The modified system is now represented as follows:
\begin{align}
\tau_L \dev{r_1}{t}{} 
&= -\left(1 + \dfrac{\tau_L}{\tau_H} \right)r_1
- \dfrac{1}{\tau_H}\int r_1dt \nonumber 
+ \beta\cos^2\left[ u_1(t) r_1(t - \tau_D) + u_2(t) \right]
+\tau_L \sigma\xi(t). \tag{6}
\end{align}
In the main text, $r_1(t)$ is expressed as $r(t)$. The default parameter values were set as follows:
$\tau_D = 0.92$ ms, $\tau_L = 15.9\ \mu$s, and $\tau_H = 1.59$ ms, 
$T_e = 3\tau_D$, and $T_c = \tau_D$.
The initial values of the control signals were
$u_1(t) = 1$ and $u_2(t) = - \pi/4$, respectively.
These parameter values are similar with the default values used in Ref.~[38] in the main text.

For the MNIST handwritten digit and Fashion-MNIST classifications,
we employed 
the time segment $\Delta t \sim 0.29337\ \mu$s.
The input images of the dataset with $28 \times 28$ pixels 
were enlarged twice
and transformed into the time series using mask pattern $\mm(t)$ defined as 
$r(t) = \mm^\tr(t) \xx$.
The elements of $\mm(t)$ were initially set randomly.
Thus, the sequence starting from the initial state is generally jagged, as shown in Fig. 2(a) in the main text. The timescale did not match the characteristic timescale of the system dynamics. After time evolution, the high-frequency components were filtered due to the finite time response.
Mask pattern $\mm(t)$ can be optimized using either the adjoint method or the DFA--adjoint method.
For the CIFAR-10, time region for the input signal was divided 
into $3 \times 32^2 = 3072$ segments using $\Delta t \sim 0.74870\ \mu$s.
The input images in CIFAR-10, comprising $32 \times 32$ RGB pixels,
were transformed into the time series as $r(t) = \mm^{\tr}(t)\xx$.

Loss function $L$ and $\GG_\mathrm{out}$ were set as the cross entropy and softmax functions for the present classification problems, respectively. 
To optimize variables $\uu(t)$, $\mm(t)$, and $\ww(t)$, we utilized the Adam optimizer with the recommended hyperparameters,
although $\alpha = 10^{-4}$ was employed to optimize $\mm(t)$.
For the training, the minibatch with a size of $100$ was utilized.

Table~\ref{tab:table1} shows 
the means and the standard variations of the classification accuracy computed by five simulations with different pseudo-random numbers for each method.
In Fig.~2(b) in the main text, we show a typical learning curve obtained by a single simulation and the averaged learning curve obtained by five simulations
as the black solid line with circles and blue solid line, respectively.
To evaluate the mean accuracy depicted in Figs.~3(b), 3(c), and 3(d), 
we used the accuracy at the last five epochs for a mean value.

The simulation code for this study was written on Python with
NumPy, an ordinary library of Python.
The training time in the DFA--adjoint method was measured as the time required for the product operation described in Eq.~(3) [or Eq.~(\ref{eq:DFA-p-timedelay})] and the update of $\uu(t)$ described in Eq.~(\ref{eq:update-u-timedelay}).
As the training time in the adjoint method, 
the time required to compute the (reverse) time evolution of $\pp(t)$
following Eq.~(2) [or Eq.~(\ref{eq_adjoint_delay_1})] 
and the update $\uu(t)$ described in Eq.~(\ref{eq:update-u-adjoint-timedelay})
was measured. 
We measured the mean training time over 20 runs, employing \verb|time()| function of \verb|time| library of Python,
where the relative standard deviations of the values were approximately 
less than 5 \% with the exception of 11.1 \% in the adjoint method with
$\tau_D = 0.92 $ ms.
Because the actual computational time depends on implementation details, numerical libraries, and programming language used, 
we should focus on the relative difference of the time in this study.

\begin{table}[htbp]
\caption{\label{tab:table1}%
Mean classification accuracy (standard deviation) obtained by five runs for MNIST, Fashion-MNIST, and CIFAR-10 datasets. 
}
\begin{ruledtabular}
\begin{tabular}{lcdr}
\textrm{Methods }&
\textrm{MNIST}&
\multicolumn{1}{c}{\textrm{Fashion-MNIST}}&
\textrm{CIFAR-10}\\
\colrule
DFA--adjoint & 0.973 (0.00075) & 0.866 (0.0074) & 0.466 (0.0086)\\
Adjoint control & 0.977 (0.0014) & 0.881(0.0012) & 0.507 (0.0075)\\
ELM (w/o control)  & 0.869 (0.0055) & 0.792 (0.0055) & 0.275 (0.0012)\\
\end{tabular}
\end{ruledtabular}
\end{table}

\begin{figure}[htbp]
\centering\includegraphics[width=8.8cm]{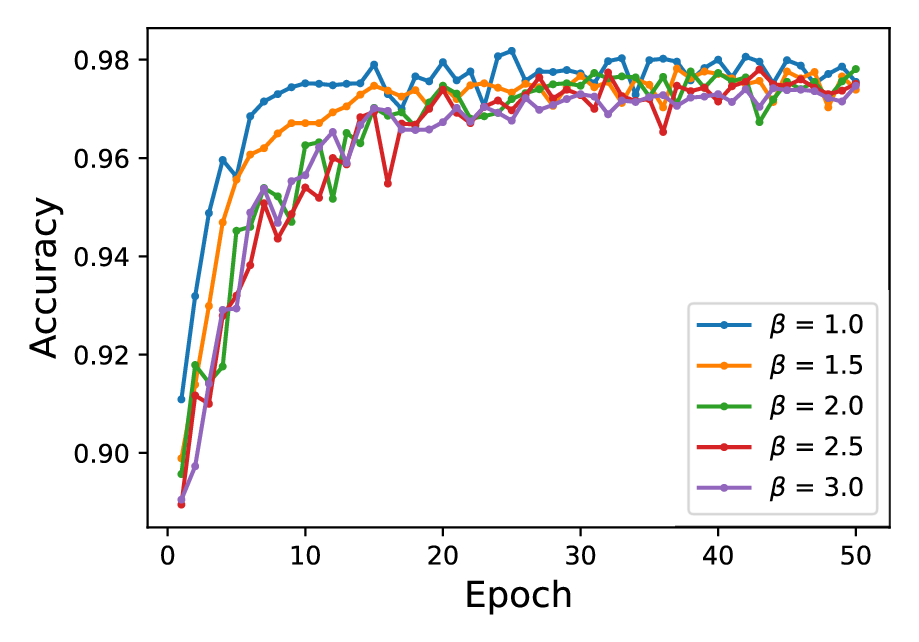}
\caption{\label{fig_s4} Numerical results of classification accuracy. The parameter values used in the numerical simulations were set as the same values as those used in the experiment.}
\end{figure}

In the numerical simulations, we used different parameter values from those used in the experiment. Even when the experimental parameter values ($\tau_D$ = 1.022 $\mu$s, $\tau_L$ = 1.59 ns, and $\tau_H$ = 530 ns) were applied in the numerical simulations, a similar classification accuracy was achieved. The results are shown in Fig.~\ref{fig_s4}. The accuracy was similar with that shown in Fig. 2(b) of the main text.

\begin{figure}[htbp]
\centering\includegraphics[width=8.8cm]{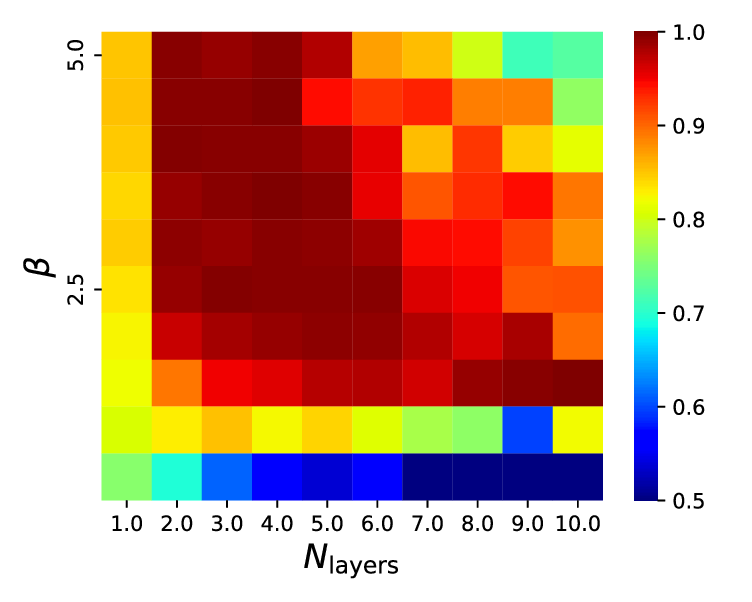}
\caption{\label{fig_twoclasses} Classification accuracy on $(\beta,N_{\mathrm{layers}})$ plane.}
\end{figure}

In the numerical results reported in the main text, $N_{\mathrm{layers}} = T_e/\tau_D$ was fixed as $3$. $N_{\mathrm{layers}}$ corresponds to the effective number of layers. We evaluated how classification accuracy depends on $N_{\mathrm{layers}}$ for a simple nonlinear two-class classification task for systematic investigations while saving computational time. The result is shown in Fig.~\ref{fig_twoclasses}, where the classification accuracy is mapped on $(\beta,N_{\mathrm{layers}})$ plane. As observed in this figure, the classification accuracy approaches 1 for $N_{\mathrm{layers}} \ge 2$ and $\beta \ge 2$. In particular, for $2 \le N_{\mathrm{layers}} \le 4$, high accuracy was obtained for a wide range of $\beta$ in this task. For a large $\beta$-values, the delay system used in this study exhibited chaotic behavior; the high sensitivity to initial states led to the instability of classification, particularly, for a long $N_{\mathrm{layers}}$. Consequently, the classification accuracy decreased. 

\section{Measurement in alignment effect}
As discussed in the main text, the update vector $\delta\uu_{DFA}$ in the DFA--adjoint method tends to align with the update vector $\delta\uu_{adj}$ in the adjoint method. To evaluate the alignment effect, we measured the cosine similarity $C$ between $\delta\uu_{DFA}$ and $\delta\uu_{adj}$, which is defined as follows:
\begin{align}
C = \dfrac{\langle
\dl\uu_{DFA}^\tr\dl\uu_{adj}
\rangle_t
}{
\lVert
\dl\uu_{DFA}
\rVert_2
\lVert
\dl\uu_{adj}
\rVert_2
},
\end{align}
where $\langle\cdot\rangle_t$ and $\lVert\cdot\rVert$ represent the time average and $L_2$ vector norm, respectively. 
The update vector $\delta\uu_{adj}$ was computed based on the trained states by the DFA--adjoint method. 
In Fig.~2(c) of the main text, we depict the mean and standard deviation of the cosine similarity at each epoch.

\section{Experiment}
The time evolution of the optoelectronic delay system used in this experiment is governed by the following equation:
\begin{align}
\tau_L \dev{r}{t}{} 
&= -\left(1 + \dfrac{\tau_L}{\tau_H} \right)r
- \dfrac{1}{\tau_H}\int rdt \nonumber \\
&+ 
\beta\cos^2\left(u(t)+\phi_{in}
\right)
\cos^2\left(
r(t-\tau_D) + \phi_{fb}
\right)
+\sigma\xi(t), \label{eq_delay}
\end{align}
where $r(t)$ denotes the light intensity (system variable).
$\tau$, $\beta$, $\tau_L$, and $\tau_H$ in the equation denote the feedback delay time, normalized gain coefficient, and time constants of the low- and high-pass filters, respectively. Specific parameter values were chosen as follows: $\tau_D = 1.022$ $\mu$s, $\tau_L = 1.59$ ns, and $\tau_H = 530$ ns. The bias points of MZM1 and MZM2 were set as $\phi_{in} = -\pi/4$ and $\phi_{fb} = - \pi/4$, respectively. 
In the experimental setup shown in Fig.~4(a), encoding the input $\xx$ as the initial state of $r(t)$ for $-\tau_D \le t \le 0$ proved challenging. Instead, we set the input data as the control signal $u(t) = \mm^{\tr}(t)\xx$ for $-\tau_D \le t < 0$.
The softmax function was used in the output layer and computed on a digital computer in this experiment, although an efficient hardware implementation can be achieved using a field programmable gate array or an application-specific integrated circuit.
The DFA--adjoint method was employed to train the control signal $u(t)$ for $t > 0$ and mask pattern $\mm(t)$ for $-\tau_D \le t \le 0$. 

\begin{figure}[htbp]
\centering\includegraphics[width=8.8cm]{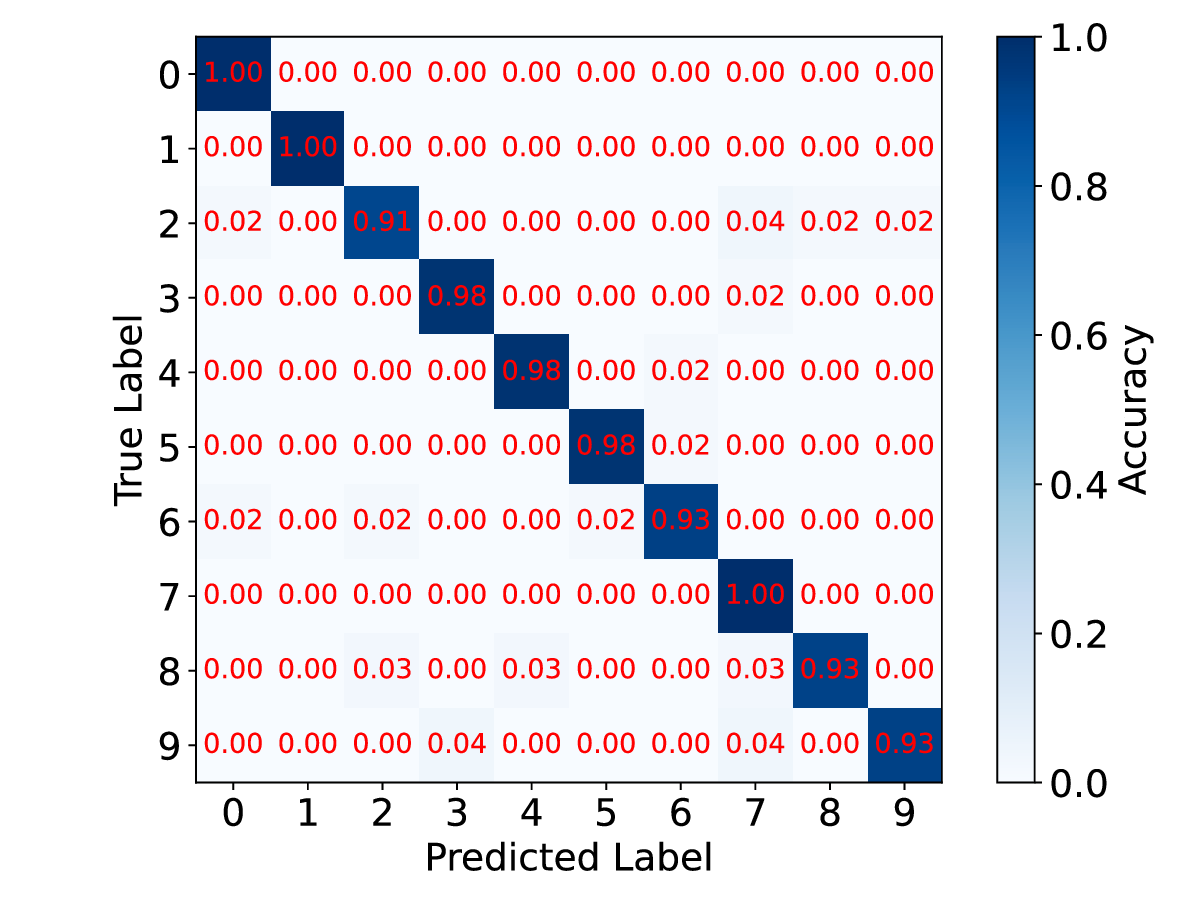}
\caption{\label{fig_confusion}
Confusion matrix. 500 test images randomly sampled from the original MNIST dataset was used for the experiment.
}
\end{figure}

For training, we randomly selected the 3000 images for the training set and 500 images for the test set from the original MNIST dataset. The distribution of the sampled images across digit labels is summarized in Table~\ref{tab:table3}. 
For the test dataset, the confusion matrix is shown in Fig.~\ref{fig_confusion}. 

\begin{table}[htbp]
\caption{\label{tab:table3}%
Number of the MNIST image samples used for training and testing.
}
\begin{ruledtabular}
\begin{tabular}{l|cccccccccc} 
\textrm{Label} & 0 & 1 & 2 & 3 & 4 & 5 & 6 & 7 & 8 & 9\\
\hline
\textrm{Training} & 285 & 339 & 299 & 295 & 325 & 273 & 306 & 329 & 261 & 288\\ 
\hline
\textrm{Test} & 42 & 67 & 55 & 45 & 55 & 50 & 43 & 49 & 40 & 54\\ 
\end{tabular}
\end{ruledtabular}
\end{table}

